\documentclass[useAMS,usenatbib]{mn2e}

\usepackage{color,graphicx}
\usepackage{aas_macros}

\begin{document}

\title[Strong lensing: image deflections from the l.o.s.]{Galaxy cluster strong lensing: image deflections from density fluctuations along the line of sight}
\author[O. Host]{Ole Host\thanks{E-mail: ohost@star.ucl.ac.uk} \\
Department of Physics \& Astronomy, University College London, Gower Street, London WC1E 6 BT, UK}
\date{8 December 2011}

\maketitle
\begin{abstract}
A standard method to study the mass distribution in galaxy clusters is through strong lensing of background galaxies in which the positions of multiple images of the same source constrain the surface mass distribution of the cluster. However, current parametrized mass models can often only reproduce the observed positions to within one or a few arcsec which is worse than the positional measurement uncertainty. One suggested explanation for this discrepancy is the additional perturbations of the path of the light ray caused by matter density fluctuations along the line of sight. We investigate this by calculating the statistical expectation value for the angular deflections caused by density fluctuations, which can be done given the matter power spectrum. We find that density fluctuations can, indeed, produce deflections of a few arcsec. We also find that the deflection angle of a particular image is expected to increase with source redshift and with the angular distance on the sky to the lens. Since the light rays of neighbouring images pass through much the same density fluctuations, it turns out that the images' expected deflection angles can be highly correlated. This implies that line-of-sight density fluctuations are a significant and possibly dominant systematic for strong lensing mass modeling and set a lower limit to how well a cluster mass model can be expected to replicate the observed image positions. We discuss how the deflections and correlations should explicitly be taken into account in the mass model fitting procedure.

\end{abstract}
\begin{keywords}
gravitational lensing: strong -- gravitational lensing: weak -- galaxies: clusters: general
\end{keywords}
\section{Introduction}
Galaxy clusters remain important tools to study the universe as a whole as well as interesting objects in their own right. Their rarity and recent formation means they are essential for understanding the formation of structure in the Universe and their high mass implies that they are bright X-ray sources and act as gravitational lenses. Both X-ray observations and weak and strong lensing can be used to infer the properties of the cluster mass distribution. $N$-body simulations of structure formation tell us that relaxed clusters should have a universal mass distribution which can be described succinctly by the NFW profile \citep*{1997ApJ...490..493N} or similar parametrizations. Since clusters formed late, they are predicted to be less centrally concentrated than the mass profile of galaxies or groups \citep{2001MNRAS.321..559B}.

Strong lensing provides information about the surface mass distribution in the innermost region of a galaxy cluster. That region is critical for measuring the inner slope and the concentration parameter of the mass profile, which can be compared to the predictions of simulations. For example, we can take a specified model for the lensing cluster mass distribution and fit the parameters of that model so as to reproduce the observed positions of the images with the least scatter. This is a common approach in the literature and there are publicly available codes such as \textsc{gravlens} \citep{2001astro.ph..2340K} and \textsc{lenstool} \citep{2007NJPh....9..447J} for that purpose. The NFW profile has been found to be a reasonable model for several galaxy clusters although there are also cases where a shallower inner slope is preferred. However, in many cases the best-fit mass model can only reproduce the observed image positions to within a few arcsec, particularly when there are several sources at a range of redshifts. Such precision is unsatisfactory given the sub-arcsec precision of the image positions in \emph{HST} imaging. For example, for the much studied cluster Abell 1689, \citet{2005ApJ...619L.143B} found a root-mean-square offset per image of 3.2 arcsec, \citet{2006MNRAS.372.1425H} found 2.7 arcsec, and \citet{2007ApJ...668..643L} reported 2.9 arcsec. Other examples are Abell 383 \citep[1.95 arcsec,][]{2011arXiv1103.5618Z}, Abell 2218 \citep[1.49 arcsec,][]{2007arXiv0710.5636E}, Abell 1703 \citep[1.4 arcsec,][]{2008A&A...489...23L}, and CL0024+1654 \citep[$\simeq2.5$ arcsec,][]{2009MNRAS.396.1985Z}.

A possible explanation for these offsets is that the path of the image light ray is not only strongly lensed by the galaxy cluster but also weakly lensed by matter density fluctuations all along the path. This means the size of the offsets will depend on the distribution and growth of structure between the source redshift and now. We refer to this process as cosmic weak lensing (CWL), to avoid confusion with the weak lensing signal from the galaxy cluster itself. We investigate the impact of CWL by calculating the expected relative image deflection angle produced by matter fluctuations from the matter power spectrum, using a formalism from lensing of the cosmic microwave background. We examine the dependence of the relative deflection on source redshift and on the image position relative to the cluster and the source. Finally we discuss how the line-of-sight matter fluctuations can be taken into account when fitting parametric mass models.

We work in the Born approximation, evaluating the lensing effects along the unperturbed line of sight and we assume flat space for simplicity. We make use of the results and notation of the CMB lensing review by \citet{2006PhR...429....1L}.  For the numerical examples we assume a standard flat $\Lambda$CDM cosmology with $h=0.71$, $\Omega_b=0.045$, $\Omega_m=0.27$ and $\Omega_\Lambda=0.73$, which is very similar to the best-fit model of the \emph{WMAP} 7-year data \citep{2011ApJS..192...18K}. We use the \textsc{iCosmo} package \citep{2011A&A...528A..33R} to calculate the evolution of cosmological quantities.

\section{Methodology}
We want to calculate the deflection angle due to CWL of an image which is strongly lensed by a galaxy cluster. Conceptually, the observed image position $\bmath\theta$ is the sum of the source position $\bmath\beta$, the strong lensing deflection caused by the cluster $\bmath{\alpha}_{\rmn{SL}}$ and the weak lensing caused by cosmic weak lensing $\bmath{\alpha}_{\rmn{CWL}}$,
\begin{equation}
\bmath{\beta}=\bmath{\theta}-\bmath{\alpha}_{\rmn{SL}}-\bmath{\alpha}_{\rmn{CWL}}
\end{equation}
We cannot expect to calculate $\bmath{\alpha}_{\rmn{CWL}}$ for a specific image. Instead, we can calculate its expectation value by treating the path of the image light ray as a random line of sight. This means that the statistics of the deflection angle are determined by the statistics of the density fluctuations, i.e.~by the 3D matter power spectrum. 

The CWL deflection angle of image $i$ is related to the lensing potential sourced by the matter along the line of sight through $\bmath{\alpha}_i=\bmath\nabla \psi_i$. The angular power spectrum of the lensing potential is $\langle \psi_i,\psi_j\rangle$, where the potential is projected from the sources at comoving distances $\chi_i$ and $\chi_j$. In the Limber approximation $(\ell \gg 1)$ the lensing potential power spectrum can be written as
\begin{equation}\label{eq:cell}
C_\ell^\psi=\frac{8\pi^2}{\ell^3}\int_0^{\chi_\star}\chi d\chi \mathcal{P}_\Psi(\ell/\chi,\chi)\left(\frac{\chi_i-\chi}{\chi_i\chi}\right)\left(\frac{\chi_j-\chi}{\chi_j\chi}\right).
\end{equation}
As usual in the Limber approximation, only the transverse modes contribute to the power spectrum so the integration extends up to the closest source, $\chi_\star=\min(\chi_i,\chi_j)$. $\mathcal{P}_\Psi(k,\chi)$ is the 3D power spectrum of the Weyl potential. This can be related to the matter power spectrum $P(k,\chi)$ through 
\begin{equation}
\mathcal{P}_\Psi(k,\chi)=\frac{9 \Omega_m^2(\chi)H^4(\chi)}{8\pi^2}\frac{P(k,\chi)}{k},
\end{equation}
where $\Omega_m$ is the matter density and $H$ is the Hubble rate.

Now we can calculate the variance of the deflection angle field from the angular power spectrum of the potential \citep[see][for details]{2006PhR...429....1L}, 
\begin{eqnarray}
C_{gl}(r) &\equiv & \langle \bmath{\alpha}_i\bmath\cdot\bmath{\alpha}_j \rangle \\
	& = & \langle \bmath{\nabla} \psi_i \bmath\cdot \bmath{\nabla} \psi_j \rangle \\
 	& = & 4\pi\int d\ell \int \chi d\chi \mathcal{P}_\Psi(\ell/\chi,\chi) F_{ij} (\chi) J_0(\ell r(\chi)),\label{eq:cgl}
\end{eqnarray}
where $r(\chi)$ is the angular separation between the light rays producing the images (which will vary with distance for strongly lensed images) and $F_{ij}(\chi)$ is shorthand for the lensing efficiency factors in equation (\ref{eq:cell}).

There can be no observable consequences if CWL causes the whole observed field to be displaced coherently. Hence, the relevant issue to consider is not how large the deflection angle $\bmath{\alpha}_i$ of image $i$ is, but rather how large is the difference of $\bmath{\alpha}_i$ and the deflection $\bmath{\alpha} '$ of a fiducial line of sight. This reference sight-line determines a `zero-point' for the CWL deflection angles in the observed field. It can be arbitrarily chosen in principle, but it is convenient to choose one that passes through the centre of the cluster lens model. The variance of the relative deflection is given by
\begin{equation}\label{eq:sii}
\sigma_{ii}^2=\left\langle (\bmath{\alpha}_i - \bmath{\alpha} ')^2\right\rangle   =  2\,C_{gl}(0)-2\,C_{gl}(r),
\end{equation}
where for both terms the integral over $\chi$ in equation (\ref{eq:cgl}) should be evaluated up to the source distance $\chi_i$. This is the expected offset of an image relative to the cluster centre. The strong lensing event causes the angular separation to vary with comoving distance,
\[
r=r(\chi) = \left\{
\begin{array}{rll}
r_0 & \rmn{for } & 0 < \chi < \chi_{\rmn{cl}}, \\	
(r_0-r_s)\frac{\chi_s-\chi}{\chi_{s}-\chi_{\rmn{cl}}} & \rmn{for }& \chi_{\rmn{cl}} < \chi < \chi_i,
\end{array} \right.
\]
and $r_s$ is the separation in the image plane between the source and the reference position. The covariance of two images of different, or identical, sources is
\begin{eqnarray}
\sigma_{ij}^2 & = & \left\langle (\bmath{\alpha}_i- \bmath{\alpha}')\bmath\cdot(\bmath{\alpha}_j-\bmath{\alpha}')\right\rangle \nonumber \\
 & = & C_{gl}(0)+C_{gl}(r_{ij})-C_{gl}(r_i)-C_{gl}(r_j), \label{eq:sij}
\end{eqnarray}
where again the angular separations will vary with comoving distance and all terms are evaluated at $\chi_\star=\min(\chi_i,\chi_j)$.

How does this relate to the cluster mass modeling? Let us summarize how parametric modeling proceeds: the best model is the one that reproduces the observed positions $\{\bmath{\beta}_i\}$ with the least amount of scatter and is typically found by minimizing a $\chi^2$-function
\begin{equation}
\chi^2=\sum_{i,j,\delta}(\beta_{i,\delta}-\mu_{i,\delta})\mathbfss{C}_{ij}^{-1}(\beta_{j,\delta}-\mu_{j,\delta}),
\end{equation}
where $\delta$ labels the RA and Dec components, the model-predicted position of image $i$ is $\bmath{\mu}_i$ and $\mathbfss{C}$ is the image position covariance. The number of degrees of freedom for $N_{\rmn{im}}$ image constraints of $N_{\rmn{so}}$ sources and a model containing $N_p$ parameters is $2(N_{\rmn{im}}-N_{\rmn{so}})-N_p$ since each observed position consists of two coordinates and the source position of each image family must be inferred as well.

In the literature, the uncertainty on the image position is most often assumed to be the positional measurement uncertainty which is of the form $\sigma_{ij}=\sigma\,\delta_{ij}$, i.e.~the covariance matrix is diagonal and the diagonal elements are the same. However, the CWL covariance matrix is more complex with diagonal elements $\sigma_{ii}^2$ and off-diagonal elements $\sigma_{ij}^2$, as given by equations (\ref{eq:sii}) and (\ref{eq:sij}). Hence, using the CWL covariance matrix will formally lead to a best-fit model which is different from the one found when only taking the positional measurement uncertainty into account.

We have stipulated that we treat the image sight-lines as random, in the sense that no other information than the source redshift and position is being used. In practice, this means that we can use the standard 3D matter power spectrum $P(k,\chi)$. A more refined treatment would also take into account the presence of the lensing cluster, which is a rare type of object located at the intersection of filaments in the cosmic structure. Hence there are likely more extreme density fluctuations in the surroundings of the cluster than what we assume. This could perhaps be modeled by biasing the input matter power spectrum in a suitably calibrated fashion. We leave this to future work.

In short, the expected deflection angle of an image relative to the cluster centre is given by $\sigma_{ii}$, equation (\ref{eq:sii}). Given a set of observed multiple image positions, the variances $\sigma_{ii}^2$ together with the covariances $\sigma_{ij}^2$ (equation (\ref{eq:sij})) form the CWL covariance matrix which should be used to find the best-fit mass model.

\section{Numerical examples}
Let us now discuss two examples to illustrate typical results. 

\begin{figure}
\begin{center}
\includegraphics[width=.9\columnwidth]{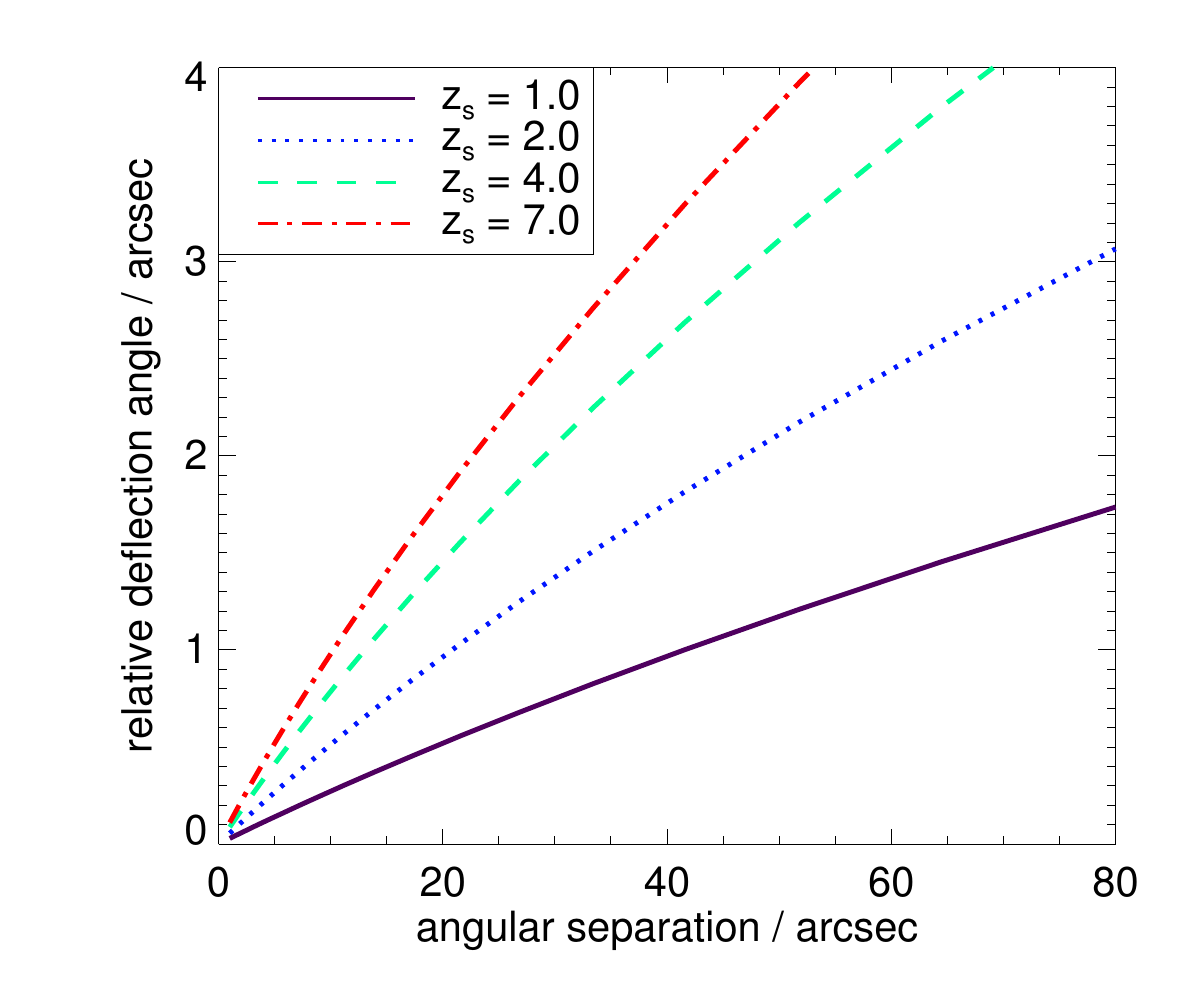}
\caption{Expectation of the relative deflection $\sigma_{ii}$ of an image as a function of distance on the sky to the cluster centre, for various source redshifts $z_s$. In this case the strong lensing event has been disregarded so that $r(\chi)=r$. Hence, this figure shows the typical scale of the relative deflection but it is not exact for a strongly lensed image.}
\label{fi:defl1}
\end{center}
\end{figure}

First we consider a single image {\bf located} some distance from the cluster centre, and for simplicity we disregard the strong lensing event, i.e.~the angular separation between the unperturbed light ray and the sight-line through the cluster is constant. Fig.~\ref{fi:defl1} shows the resulting expected relative deflection $\sigma_{ii}$ and how it increases with the separation and source redshift. It is clear that CWL deflections can easily produce deflections of a few arcsec, increasing with separation on the sky, and also that a large fraction of the total deflection is contributed from evolved structure in the late universe. Fig.~\ref{fi:defl1} can be used to estimate the scale of the relative deflection for a given redshift, but for strongly lensed images the specific $r(\chi)$ of the path of the light ray must be taken into account for a precise result.

\begin{figure}
\begin{center}
\includegraphics[width=.9\columnwidth]{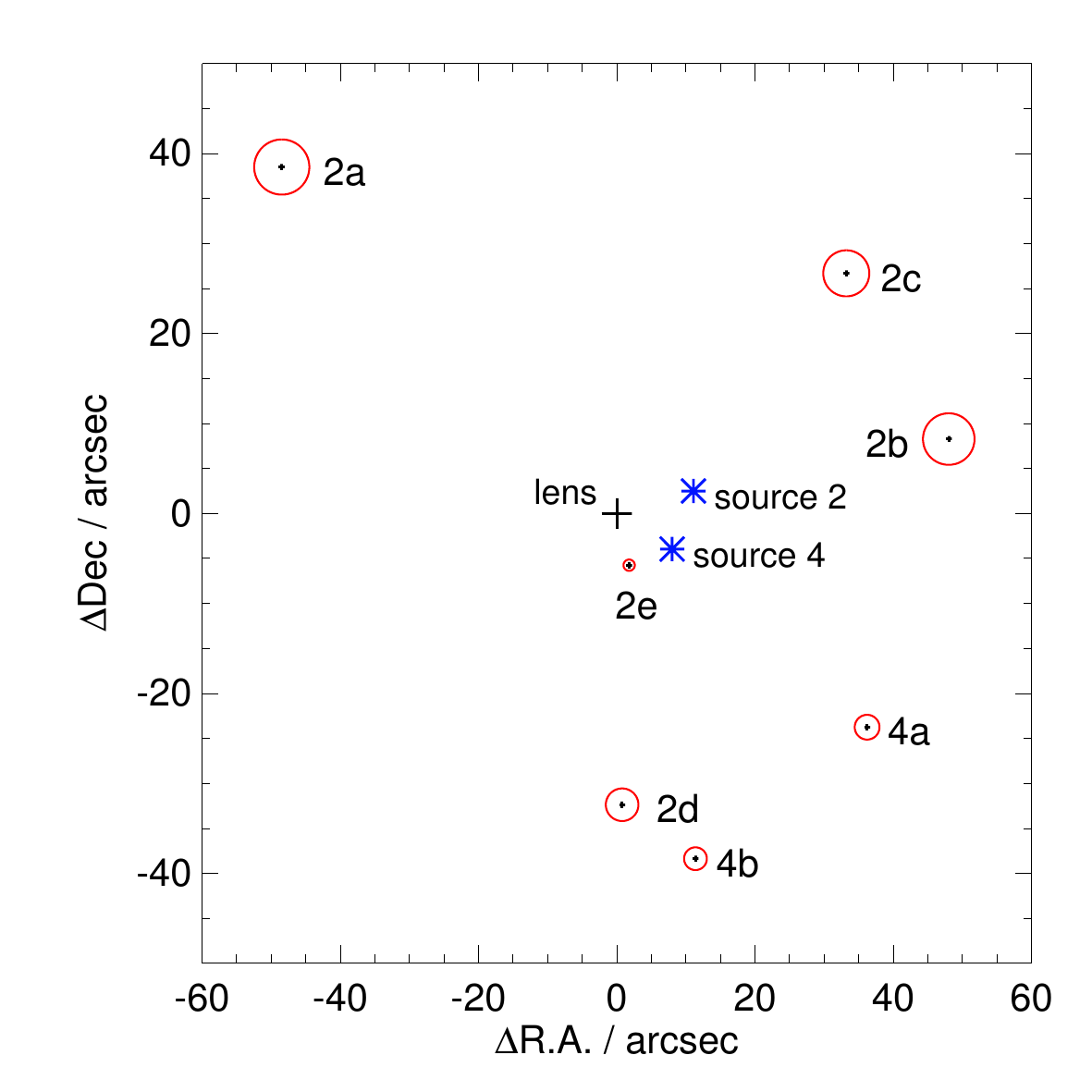}
\caption{Schematic representation of the central $2\times2$ arcmin of the Abell 1689 field showing the five images (a-e) of the background source 2 at $z=2.5$ (following the usual labeling in the literature) as well as two of the five images of source 4 ($z=1.1$). The centre of the cluster is marked with a '+'. The red circles show the $2\sigma_{ii}$ error circles as given by equation (\ref{eq:sii}), assuming that the true position of the sources are at the marked positions. Table \ref{tb:1689} gives the numerical values as well as the image correlation matrix.}
\label{fi:1689}
\end{center}
\end{figure}

The second example is Abell 1689 (e.g.~\citet{2005ApJ...619L.143B,2007ApJ...668..643L}) which has a very large number of multiple images. We consider a small subset of these to illustrate the calculations. Source 2 at $z=2.5$ has five images located between 6 and 60 arcsec from the centre of the cluster while source 4 at $z=1.1$ also has five images, of which we consider two that are both $\sim40$ arcsec from the centre. Fig.~\ref{fi:1689} shows the structure of the field. We calculate all the elements of the covariance matrix $\sigma_{ij}^2$, taking the varying separations $r(\chi)$ into account by assuming reasonable choices for the true source positions. The expected relative deflection angles $\sigma_{ii}$ of these images w.r.t.~the cluster centre are given in Table \ref{tb:1689} and shown in Fig.~\ref{fi:1689} as error circles. Again, it is clear that the impact of CWL increases with the angular separation of the images and with the redshift of the sources. E.g., images 4a and 4b have expected relative deflections which are half as large as that of image 2c, even though all three images are about equally far from the cluster. Another case is image 2e which has a much smaller expected deflection than 2a--2e since it is very close to the cluster centre. In \citet{2007ApJ...668..643L}, the rms image plane offset per image for system 2 was $1.62$ arcsec. Taking a naive rms average of the five $\sigma_{ii}$ (and hence disregarding correlations) we obtain an expected offset per image of $1.52$ arcsec, i.e.~a very similar number. The correlation matrix (see Table \ref{tb:1689}) shows various degrees of correlation between the deflections of the images, e.g. it can be seen that images 2b and 2c which are relatively close together have strongly correlated deflections with a correlation coefficient of 0.88. This is of course not surprising since the light rays producing each image must have passed through much the same density fluctuations. On the other hand, images 2d and 4b are even closer together but have a slightly smaller correlation coefficient of 0.82 due to the lower redshift of source 4.

\begin{table*}
\begin{minipage}{130mm}
\caption{Multiple images lensed by Abell 1689: source redshift, observed distance on the sky to the cluster centre, expected relative deflection $\sigma_{ii}$, and correlation matrix $\frac{\sigma_{ij}^2}{\sigma_{ii}\sigma_{jj}}$.}\label{tb:1689}
\begin{tabular}{lrrrrrrrrrrr}\hline
ID & $z$ & $r$ / arcsec & $\sigma_{ii}$ / arcsec & Corr \\\hline
2a & $2.5$ & $61.9$ & $1.97$& $1.00$ \\
2b & \ldots & $48.7$& $1.84$& $-0.29$ & $1.00$\\
2c & \ldots & $42.6$& $1.65$& $-0.03$ & $0.88$ & $1.00$\\
2d & \ldots & $32.4$& $1.17$& $-0.45$ & $0.11$ & $-0.20$ &$1.00$ \\
2e & \ldots & $ 6.0$& $0.40$& $-0.48$ & $0.53$ & $0.32$ & $0.60$ & $1.00$&\\
4a & $1.1$ & $43.3$ & $0.89$ & $-0.55$ & $0.59$ & $0.30$ & $0.53$ & $0.48$ & $1.00$ \\
4b & \ldots & $40.0$ & $0.82$ & $-0.48$ & $0.20$ & $-0.10$ & $0.82$ & $0.49$ & $0.78$ & $1.00$ \\
\hline 
\end{tabular}
\end{minipage}
\end{table*}

\section{Applications}
While CWL can explain at least a significant part of the offsets between best-fit model prediction and observed image position, it should really be taken into account already at the point of model fitting. As discussed above, the covariance matrix $\sigma_{ij}^2$ can be inserted straight-forwardly into existing methods based on $\chi^2$-minimization in the image plane. In practical strong lensing applications, it is not necessarily feasible to treat the source positions as free parameters to be marginalized over. Instead, the family of images are projected backwards and the mean of these images is treated as a proxy for the source position which can then be projected forwards to the image plane. While this is not probabilistically rigorous, it reduces the computational load significantly. It should be relatively straight-forward to adapt our treatment of LSS to such a framework. Each backwards projection could be weighted by $\sigma_{ij}^2$, the resulting weighted mean would be the assumed source position, and the rest of the analysis could proceed as discussed above. There are also methods in which the $\chi^2$-function is defined in the source plane, which is computationally even more efficient but suffers from possible biases. It appears to be more of a challenge to use our method in such a case, since the covariance matrix $\sigma_{ij}^2$ would need to be transformed back to the source plane while taking into account both shear and magnification effects. 

\begin{figure}
\begin{center}
\includegraphics[width=.9\columnwidth]{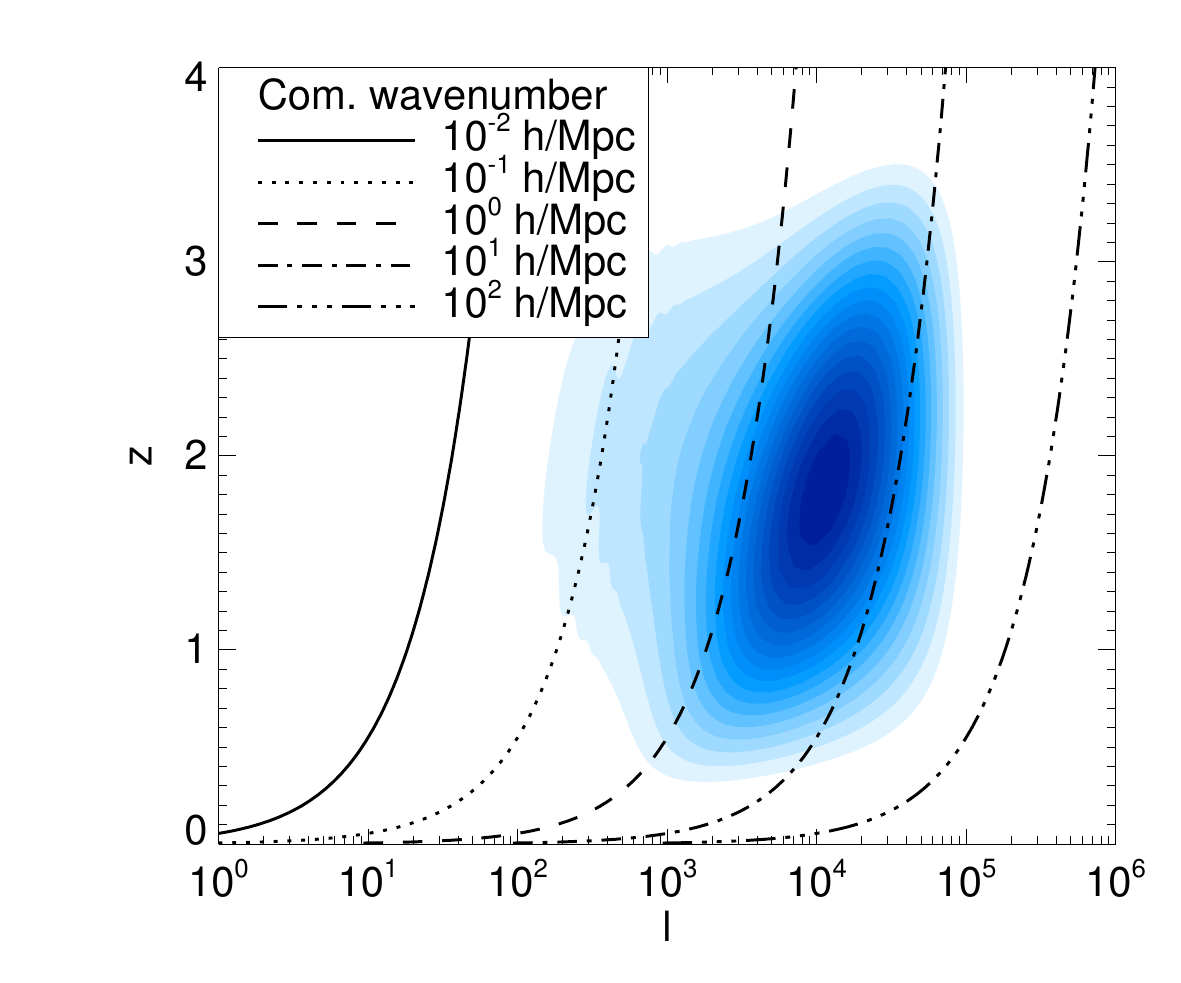}
\caption{Where does the relative deflection angle signal come from? The contours show the integrand of $\sigma_{ii}^2=2C_{gl}(0)-2C_{gl}(r)$ for $r=10$ arcsec (see equation(\ref{eq:cgl})). The lines show the co-moving wavenumber $k$ of the matter power spectrum. Most of the signal contributing to $\sigma_{ii}^2$ comes from the region between $1$ and $10\,$h/Mpc. For smaller $r$, the signal region will move to the right.}
\label{fi:conv}
\end{center}
\end{figure}

A different issue relates to the small size of strong lensing fields which is between some tens of arcsec across and one or two arcmin. This means we are predicting CWL effects on extremely small scales and we must worry about our ability to calculate $P(k,\chi)$. Fig.~\ref{fi:conv} displays the integrand of $C_{gl}(0)-C_{gl}(r)$ and it can be seen that most of the signal comes from the region of parameter space where $k=1-10\,h/\rmn{Mpc}$. This is far into the nonlinear region where baryonic effects also play a very significant role. The results discussed above were calculated using the popular halo model prescription of \citet{2003MNRAS.341.1311S} which is calibrated on dark matter-only $N$-body simulations and cannot be expected to deliver precise power spectra for very high $k$. There are simulations of  structure formation including baryonic processes which are improving rapidly but there does not appear to be consensus regarding the various feedback processes that play an important role at the moment. To estimate the possible impact of this issue we have used the recent simulation results of \citet{2011MNRAS.415.3649V}, who measured the small scale matter power spectrum using the OWLS suite of simulations \citep{2010MNRAS.402.1536S}. Their `most realistic' simulation based on the \emph{WMAP7} cosmology follows the formation of structure down to scales of $\sim 500 h/\rmn{Mpc}$ and it includes AGN feedback and does not suffer from the over-cooling problem. We take the matter power spectrum, which has about twice as much power at $k=100\,h/\rmn{Mpc}$, from that simulation for the range $k>1\,h/\rmn{Mpc}$ for all redshifts as an input and test how much the example results change. Comparing the case of parallel light-rays (Fig.~\ref{fi:defl1}), for a source redshift of $z_s=1$ the relative deflection angle $\sigma_{ii}$ increases by 9\% for a separation of $r=3$ arcsec and by 2\% for $r=30$ arcsec, while for a source redshift of $z_s=7$ the corresponding increases are 11\% and 1\%, respectively. Hence, the effect of the changed matter power spectrum is relatively benign for large angular separations while for small angular scales the difference is smaller than a tenth of an arcsec, and hence less than the pixel size in \emph{HST} imaging. Hence, we conclude that uncertainty about the small scale power spectrum is a minor complication at present.

\section{Discussion} 
Finally, we compare our analysis with the literature and summarise.

\citet{2011MNRAS.411.1628D} investigated the effect of CWL on the positions of strongly lensed images by ray-tracing through multiple lens planes, populated with NFW halos from the Millennium $N$-body simulation (MS, \citet{2006Natur.440.1137S}). They found that the mean displacement of the light-rays was $9\,$kpc (comoving), corresponding to 2.3 arcsec at $z=0.2$. Since they average over a distribution of source redshifts it is difficult to make a direct comparison with our results, but we do note that their results are comparable in order of magnitude. They also find that structure associated with the cluster lens ('correlated large-scale structure') can perturb the image positions, but to a lesser degree than CWL. In a similar analysis also based on the MS, \citet{2010Sci...329..924J} found that CWL added deflections of typically 1 arcsec while correlated large-scale structure contributed 0.6 arcsec. They also noted that, in a few cases, CWL could change the multiplicity of images, an effect which we do not take into account here. 

Unlike for the strong lensing regime, the impact of CWL on the weak lensing signal measured further away from the centre of a cluster lens has been studied in detail, e.g.~\citet{,2003MNRAS.339.1155H} or \citet{2004PhRvD..70b3008D}. Recently, \citet{2011ApJ...740...25B} and \citet{2011MNRAS.412.2095H} confirmed in simulations that CWL is an irreducible systematic to WL cluster mass measurements, but that it can be accounted for stochastically, i.e.~in a similar spirit to the present work (see also \citet{2011MNRAS.416.1392G}). In \citet{2011ApJ...738...41U} and \citet{2011arXiv1103.5618Z}, the CWL contribution to the measured surface mass density in radial bins was taken into account for a joint strong lensing-weak lensing analysis, using the predicted bin-to-bin covariance calculated from cosmic shear power spectra.

Another proposed explanation for the positional offsets is substructure in the lensing cluster. Individual cluster galaxies can be modelled concisely but coarsely through scaling relations, and \citet{2010Sci...329..924J} found that a 20\% random variation in the scaling relations could introduce scatter in the image positions of up to 1 arcsec. This appears to be slightly less significant than our prediction for the effect of CWL, but we do not claim that CWL is the dominant source of image position uncertainty, only that it is a likely explanation for the reported offsets.

In summary, we have presented a reasonable explanation for the offsets between the observed and the model-predicted image positions which are often encountered in current parametric modeling of strong lensing galaxy clusters. We find that additional lensing by density fluctuations along the line of sight -- cosmic weak lensing -- can displace images by a few arcsec, which is similar to the discrepancies reported in the literature. Since CWL is an unavoidable and possibly dominant systematic, we propose to account for this by fitting the mass model according to the image position covariance matrix of CWL. We intend to do this for the CLASH sample \citep{2011arXiv1106.3328P} of 25 galaxy clusters, which are currently being observed. This will allow us to infer parametric mass models with much better understanding of the uncertainties on the model parameters.

\section*{acknowledgments}
We thank Sarah Bridle and Ofer Lahav for discussions and Adi Zitrin for comments and data.

\bibliographystyle{mn2e}
\bibliography{betacluster}

\end{document}